\documentstyle[aps,prl,twocolumn,graphicx]{revtex}
\begin{document}
\wideabs{
\title{Anomalous NMR Spin-Lattice Relaxation in SrB$_{6}$ and
Ca$_{1-x}$La$_{x}$B$_{6}$}

\author{J. L. Gavilano,$^{1}$ Sh. Mushkolaj,$^{1}$ D. Rau,$^{1}$
H. R. Ott,$^{1}$ A. Bianchi,$^{2}$ D. P. Young,$^{2}$ and Z. Fisk$^{2}$}

\address{
$^{1}$ Laboratorium f\"{u}r Festk\"{o}rperphysik, 
ETH-H\"{o}nggerberg, CH-8093~Z\"{u}rich, Switzerland \\
$^{2}$National Magnetic Field Laboratory, Florida State University, 
Tallahassee, Florida 32306
}


\maketitle 

\begin{abstract}
We report the results of $^{11}$B nuclear magnetic resonance (NMR)
measurements of SrB$_{6}$ and Ca$_{0.995}$La$_{0.005}$B$_{6}$ below
room temperature.  Although the electrical resistivities of these two
materials differ substantially, their $^{11}$B-NMR responses exhibit
some strikingly common features.  Both materials exhibit ferromagnetic
order, but their $^{11}$B-NMR spectra reveal very small hyperfine
fields at the Boron sites.  The spin lattice relaxation $T_{1}^{-1}$
varies considerably with external field but changes with temperature
only below a few K. We discuss these unusual results by considering 
various different scenarios for the electronic structure of these 
materials.
\end{abstract}
\vspace*{-0.9cm}

} 

   Hexaboride compounds XB$_{6}$ (X= elements of the alkaline-earth
   and lanthanide series) have previously been studied
   \cite{Etourneau85}, but they continue to reveal new and unexpected
   physical properties~\cite{Ott98}.  This is particularly true for
   the case of divalent hexaborides such as
   EuB$_{6}$~\cite{Digiorgi97,Gavilano98,Ambrosini99},
   CaB$_{6}$~\cite{Vonlanthen2000} and SrB$_{6}$~\cite{Ott97} whose
   physical properties seem to depend on subtle details of the
   electronic band structure, and are very sensitive to the presence
   of microscopic
   imperfections~\cite{Vonlanthen2000,Massida97,Rodriguez2000}.  
   Recent research efforts involving these systems resulted in an
   unexpected discovery of ferromagnetic ordering of tiny magnetic
   moments in La-doped CaB$_{6}$~\cite{Young99} with Curie
   temperatures of the order of 600 K.  Subsequently,
   ferromagnetic order was also encountered in the nominally binary
   compounds SrB$_{6}$ and CaB$_{6}$, as well as La-doped
   BaB$_{6}$~\cite{Ott98,Vonlanthen2000,Ott2000,Terashima00}.

In this letter we present and discuss NMR data obtained on
ferromagnetic SrB$_{6}$ and Ca$_{0.995}$La$_{0.005}$B$_{6}$.  The
compounds XB$_{6}$ all crystallize by adopting the CaB$_{6}$ type
structure with the space group \textit{Pm3m}.  The Boron ions,
directly probed by our NMR experiments, form a rigid network of
interconnected Boron octahedra.  The point symmetry of the B-sites is
\textit{4mm} and allows for nonzero axially-symmetric electric-field
gradients.  Somewhat unexpectedly,  the
$^{11}$B-NMR response is largely insensitive to the magnitude of the
electrical resistivities of these materials.  At 1 K, the value of the
electrical resistivity $\rho$ of the two samples differs by a factor
of 30, but there is no appreciable difference in the spin-lattice
relaxation rate $T_{1}^{-1}$ at this temperature.  The $T_{1}^{-1}(T)$
data reflect neither a metallic nor a semiconducting behavior. 
Furthermore the data do not provide straightforward evidence for
the ferromagnetism that is indicated by $M(H)$ measurements on the
same samples\cite{Young99,Ott2000}.  However, the spin-lattice
relaxation rate is orders of magnitude larger than what may be expected
from a relaxation via itinerant electrons alone.  We interpret this as
evidence for the presence of weakly localized electronic states, with
 unusual dynamics.  These states coexist with extended states
responsible for the electrical transport and may be related to the
unusual ferromagnetism observed in these materials.

\begin{figure}[t]
\begin{center}\leavevmode
\includegraphics[width=0.8\linewidth]{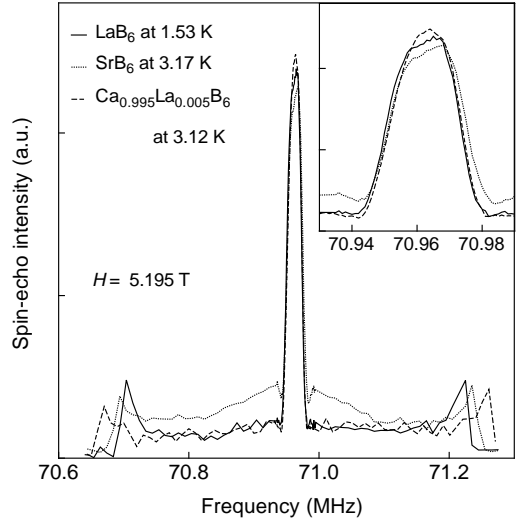}
\caption{ 
$^{11}$B-NMR spectra for SrB$_{6}$, Ca$_{0.995}$La$_{0.005}$B$_{6}$ 
and LaB$_{6}$ measured at a fixed applied field of 5.195 T and at temperatures of 3.17, 3.12 and 
1.53 K, respectively.
} 
\protect 
\label{Figure1}
\end{center}
\end{figure}
For our NMR experiments we used standard spin-echo techniques.  The
samples consisted of collections of small single crystals, grown by
the Al-flux method.  The materials have extensively been characterized
by measurements of thermal and transport
properties\cite{Vonlanthen2000,Ott97,Young99}.

In Fig. \ref {Figure1} we display examples of the recorded $^{11}$B
NMR spectra for SrB$_{6}$, Ca$_{0.995}$La$_{0.005}$B$_{6}$ and
LaB$_{6}$, measured at a fixed applied field of 5.195 T and at
temperatures of 3.17, 3.12 and 1.53 K respectively.  The shape of the
$^{11}$B-NMR spectra is characteristic for nuclear spins $ I = 3/2$ in
powder samples with randomly oriented grains.  From the extension of
the wings of the spectra we infer quadrupolar frequencies, $\nu_{Q} =
e^{2}qQ/2h$, of 600, 550 and 530 kHz for
Ca$_{0.995}$La$_{0.005}$B$_{6}$, SrB$_{6}$ and LaB$_{6}$ respectively. 
The corresponding electric field-gradients at the B sites are 1.40,
1.28 and $1.23 \cdot 10^{21}$V/m$^{2}$, respectively, in excellent
agreement (better than 5 \%) with theoretical values predicted for
SrB$_{6}$ and CaB$_{6}$~\cite{Schwarz96}.

The prominent signal near the center of the spectra represents the
+1/2 to -1/2 $^{11}$B-nuclear Zeeman transition, and its position
indicates that the NMR lineshift is very small, of the order of 80 ppm
or less.  Below room temperature we found no substantial
temperature-induced changes in the position of the NMR line for any of
these materials.  The width of the central line, FWHM = 23 kHz, is of
the order of those of good metallic samples.  No appreciable
differences are found in the position and width of the central
transition for LaB$_{6}$ with 1 conduction electron per unit cell and
SrB$_{6}$ as well as Ca$_{0.995}$La$_{0.005}$B$_{6}$, with conduction
electron concentrations $n_{c} \leq 5 \cdot 10^{-3}$ electrons per
unit cell and exhibiting ferromagnetic ordering below 600 K.
In ferromagnetic powders the NMR linewidth is largely determined by
demagnetization factors, yielding a variation of the local fields of
the order of the magnetization.  In our case this implies a magnetic
moment of 10$^{-1}$ Bohr magnetons per unit cell, an upper limit
consistent with our data.  The actual ordered moment may be much
smaller.
\begin{figure}[t]
\begin{center}\leavevmode
\includegraphics[width=0.8\linewidth]{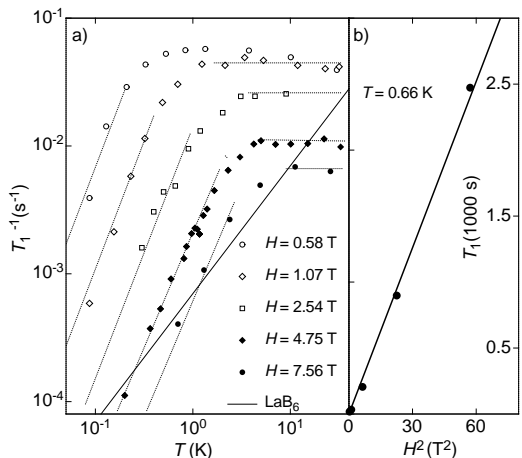}
\caption{ (a) $T_{1}^{-1}(T)$ for SrB$_{6}$ for various applied
magnetic fields.  The dotted lines are guides to the eye.  The solid
line represents $T_{1}^{-1}(T)$ for LaB$_{6}$.  (b) $T_{1}$ for
SrB$_{6}$ as a function of the applied field at $T$ = 0.66 K. }
\protect
\label{Figure2}
\end{center}
\end{figure}

In Fig.  2a we display the temperature dependence of the spin-lattice
relaxation rate $T_{1}^{-1}$ of SrB$_{6}$ for various applied magnetic
fields.  The $T_{1}^{-1}$'s were obtained from one-parameter fits to
the nuclear magnetization recovery $m(t)$ of the central transition
after it was completely destroyed by the application of a short or a
long sequence of $rf$-pulses.  The $m(T)$ data at the wings of the
signal also follow the expected theoretical function (different from
the one for the central line) for the case of magnetic relaxation and
yield the same $T_{1}$ values.  Two different regimes above and below
a field-dependent crossover temperature $T_{0}$ may be identified.  At
low temperatures, $T < T_{0}$, $T_{1}^{-1} \propto
1/(H^{2}T^{\alpha})$ with $\alpha$ of the order of -2 (see Figs.  2a
and 2b), and at high temperatures $T_{1}^{-1}$ is approximately
$T$-independent and proportional to $1/H$ for $H > 1$ T. The strong
field dependence of $T_{1}^{-1}(T)$ also suggests that the relaxation is of
magnetic origin, and therefore invokes electronic degrees of freedom. 
Note that $T_{1}^{-1}(T)$ cannot be reconciled with the expectations
for a conventional ferromagnet.  For $T \ll T_{C}$, $i.e.$, below 50 K
and ferromagnetic order among well localized moments, $T_{1}^{-1}(T)$
is expected to vary as a power-law in $T$ or
exponentially-in-$1/T$\cite{Beeman68}.  In case of ferromagnetic order
among small and itinerant magnetic moments, $T_{1}^{-1}(T) \propto
T$\cite{Moriya79}.  Neither case agrees with our data, therefore, we
have to consider other sources of spin-lattice relaxation.

We first consider a relaxation via spurious paramagnetic
impurities, hereafter referred to as impurity relaxation.  In the case
of impurity relaxation, $T_{1}^{-1}$ usually scales linearly with the
impurity concentration $n_{imp}$ for a large range of values, as
observed, $e.g.$, for Ga impurities in LaAl$_{2}$\cite{McHenry72}
where $T_{1}^{-1} \propto n_{imp}$ for $n_{imp}$ between 0.001 and 0.1. 
Since $T_{1}^{-1}(T)$ for SrB$_{6}$ and
Ca$_{0.995}$La$_{0.005}$B$_{6}$ coincide (see Fig.  3), the assumption
of a substantial impurity relaxation would imply that both materials
would suffer from, within a few percent, the same concentration of the
same type of paramagnetic impurities, a rather unlikely situation. 
Further evidence against this scenario is provided by the results of
S. Kunii\cite{Kunii99} who, using ESR techniques, searched for but did
not find hints for magnetic impurities, such as Fe and Eu, in a
ferromagnetic sample of Ca$_{0.995}$La$_{0.005}$B$_{6}$.  We conclude
that $T_{1}^{-1}$ in SrB$_{6}$ and Ca$_{0.995}$La$_{0.005}$B$_{6}$ is
not dominated by unintended paramagnetic impurities.
\begin{figure}[t]
\begin{center}\leavevmode
\includegraphics[width=0.8\linewidth]{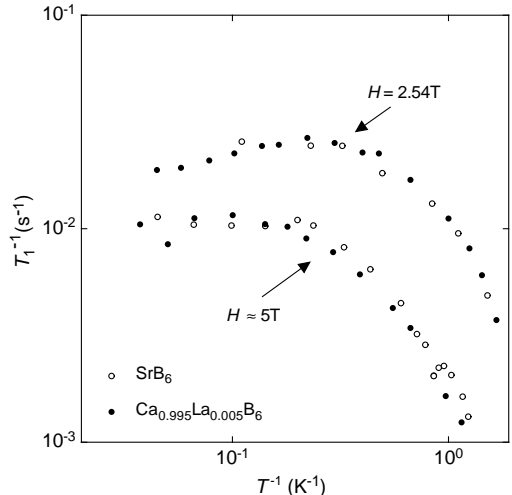}
\caption{ $T_{1}^{-1}$ as a function of $1/T$ for SrB$_{6}$ and Ca$_{0.995}$La$_{0.005}$B$_{6}$
for two different applied magnetic fields.  Note the striking
coincidence of the relaxation for the different
materials.} 
\protect
\label{Figure3}
\end{center}
\end{figure}

Next we consider the relaxation due to conduction electrons.  It has
been found that for LaB$_{6}$, above 30 K, $(T_{1}T)^{-1} \approx
7\cdot10^{-4}$ K$^{-1}$s$^{-1}$~\cite{Gavilano98,Ambrosini99}.  This
small value is consistent with the observed NMR lineshift and
indicates a very weak coupling between the conduction electrons and
the B nuclei in LaB$_{6}$.  From a simple scaling of $(T_{1}T)^{-1}$
by considering the differences in $n_{c}$, a three orders of magnitude
weaker spin-lattice relaxation is expected for SrB$_{6}$ and La-doped
CaB$_{6}$.  However, as shown in Fig.\enspace2, $(T_{1}T)^{-1}$ of
SrB$_{6}$ below 30 K is larger than or equal to the values of
LaB$_{6}$ extrapolated from the results of the high-temperature
measurements at 4.7 T~\cite{Gavilano98,Ambrosini99}.  Thus the direct
relaxation rate $T_{1}^{-1}$ due to conduction electrons is too small
to be of relevance for the alkaline-earth hexaborides.  Naturally,
this poses the problem of identifying a source, other than the
low-energy excitations of conduction electrons to account for the
spin-lattice relaxation data.  Below we consider three different
scenarios that have recently been discussed in connection with the
electronic structure of these
materials\cite{Massida97,Tromp2000,Zhitomirsky99,Gloor2000}.

First we consider the case of a band overlap at the $X$ point of the
Brillouin zone for divalent hexaborides.  For this case, as
pointed out by Zhitomirsky and Rice \cite{Zhitomirsky99}, a Bose
condensation of electron-hole pairs, excitons, is likely to occur,
leading to ferromagnetic order among tiny moments below a high Curie
temperature $T_{C}$ upon doping.  Little is known about the dynamic
susceptibility of the excitonic condensate.  On very general grounds,
however, and drawing parallels with the case of superconductivity, at
very low temperatures ($T \ll T_{C}$) one may expect a very small, and rapidly decreasing
relaxation rate upon decreasing temperature, which is not observed here
for $T_{0} \leq T \ll T_{C}$. 

Next we consider the case of a very small gap at the $X$ point of the
Brillouin zone, $i.e.$, a gap of the order of the excitonic
binding energy.  The excitonic condensate mentioned in the previous
scenario would not form, but for this case a novel spin-lattice
relaxation mechanism has recently been suggested\cite{Gloor2000}. 
Under certain conditions the formation of ``multiexciton molecules''
containing several electron-hole pairs in SrB$_{6}$ and CaB$_{6}$ near
impurities seems energetically favorable.  The exciton molecules are
expected to carry an uncompensated spin 1/2 and their
thermally-induced spin flips are very effective in producing nuclear
relaxation.  The technical details of this interesting picture, in
particular the temperature dependence of the spin-lattice relaxation,
still have to be worked out.

Since a temperature independent $T_{1}^{-1}$ is likely to be caused by
localized magnetic moments, we finally consider yet another case. 
With a gap at the $X$ point of the Brillouin zone that exceeds the
excitonic binding-energy\cite{Tromp2000}, the conditions for the two
previous cases are not fulfilled.  Instead we consider a
relaxation due to localized electronic states, as they occur
in doped semiconductors close to the metal-insulator transition.  
Here the NMR relaxation is caused by thermally-induced
spin-flips of weakly bound holes around vacancies on the
alkaline-earth sites, and electrons below a certain mobility edge.  We
refer to both of them as weakly localized electronic states. Since
the physical picture is relatively simple, we first check the
consistency of such a scenario.

Weakly localized electronic states contribute very little to the
electronic transport at low temperatures but, contrary to the extended
electronic states may, as discussed below, cause a temperature
independent relaxation.  Mobile electrons and holes formed by
conduction band states of Ca and Sr and by $p$-orbitals of B,
respectively \cite{Massida97} are, via the magnetic dipolar
interaction, only weakly coupled to the B-nuclei.  For a weakly
localized state with a wave-function effective radius $r$, however,
the dipolar field at the B sites within the orbital is proportional to
$r^{-3}$.  Its hyperfine coupling to the B-nuclei increases rapidly as
$r$ decreases.  In doped semiconductors near the IMT, weakly localized
and extended states coexist~\cite{Fuller96}, and we assume the same to
occur for SrB$_{6}$ and Ca$_{0.995}$La$_{0.005}$B$_{6}$.  The extended
electronic states induced by defects and doping contribute to the
electronic transport and are only weakly coupled to the localized
electrons.  If localized states cause the experimental observations,
they should have approximately the same concentration for various
divalent hexaborides, near the critical concentration for
delocalization.  Judging from the $\rho(T)$ data these states are of
the order of 10 to 20 meV below the delocalization
threshold~\cite{Vonlanthen2000,Ott97}.  Consider, for example, a
single hole bound to a vacancy with an effective charge $-2e$, at a
Sr site in SrB$_{6}$.  The effective mass of the
holes~\cite{Massida97,Rodriguez2000} $|m^{*}| = 0.24$ $m_{e}$, with
$m_{e}$ as the free electron mass, and the dielectric constant
$\epsilon \approx 6$~\cite{Rodriguez2000,Degiorgi2000}, yield for the
$H-$type state an effective radius $r_{0}$ $\approx 9$ lattice
constants.  Delocalization ($i.e.$, the IMT) is expected for a
concentration $N_{i} \approx r_{0}^{-3} \approx 2 \cdot 10^{19}$
cm$^{-3}$.  The binding energy of a single $H$-type state $E_{b}
\approx 0.18$ eV is about one order of magnitude too large to match
the $\rho(T)$ data, but the energy needed for delocalization rapidly
decreases from $E_{b}$ as the IMT is approached from the insulating
side.  Of course the relatively large value of $N_{i}$ should have
measurable effects on the thermal and transport properties.  This
seems to be the case for SrB$_{6}$, and
Ca$_{0.995}$La$_{0.005}$B$_{6}$, where anomalies in the temperature
dependence of the specific heat have been found at low temperatures. 
For SrB$_{6}$, $e.g.$, the entropy associated with the excess specific
heat below 7K saturates at 0.05$\%$ of $R\cdot\ln2$ at 7 K. For
spin-1/2 particles this implies a concentration of $0.7 \cdot 10^{19}$
cm$^{-3}$, a value close to $N_{i} \approx 10^{19}$ estimated above.

Following these arguments the $T_{1}^{-1}$ data for $T > T_{0}$ may
qualitatively be understood.  The NMR relaxation rate caused by
localized electronic states interacting with the nuclei via dipolar
coupling is given by\cite{Hoch88} 
\begin{equation}
	\frac{1}{T_{1}} \approx 
	\frac{2}{5}\frac{ \gamma_{S}^{2} \gamma_{I}^{2} \hbar^{2} 
	S(S+1)}{r_{0}^{6}} \cdot f(\omega) = C\cdot f(\omega) \; ,
	\label{eq:1}
\end{equation}
 where $\gamma_{S}$ and $\gamma_{I}$ are the gyromagnetic ratios for
  electrons and $^{11}$B nuclei, respectively, $S$ is the electron
 spin and $f(\omega)$ is the spectral function describing the thermal
 fluctuations of the localized-electron system.  The Larmor frequency
 $\omega$ is proportional to the applied magnetic field.  Most of the
 spectral weight of $f(\omega)$ is assumed to be contained in a finite
 range of frequencies, from 0 up to $1/\tau$, where $\tau$ is the
 (electron) spin correlation time.  Since it is not possible to
 calculate $\tau$ from first principles, except for very simple
 situations, one usually assumes the high-temperature approximation
 $f(\omega) = \tau/(1 + \omega^{2}\tau^{2})$.

 To keep the arguments simple we neglect here complications due to spin
 diffusion\cite{Spindiffusion}.  Following the ideas of Gan and
 Lee\cite{Gan86}, the interaction between localized electrons near the
 IMT results in the formation of pairs with a broad distribution of
 singlet-triplet splittings $J$.  When $J - 2 \mu_{e}H \approx 2
 \mu_{n}H$ the random singlet-triplet transitions produce nuclear
 relaxation very effectively.  The localized electron system is then
 characterized by a broad distribution $P$ of correlation times
 $\tau$.  The model is completed with
\begin{equation}
		f(\omega) = \int_{\tau_{min}}^{\tau_{max}} P(\tau) \frac{\tau}{ 1 + 
		\omega^{2}\tau^{2} }, \qquad \omega\tau_{min} \ll 1 \ll \omega\tau_{max}.
\label{eq:2}
\end{equation}
Since in our case $P(\tau) \propto 1/\tau$\cite{Hoch88},  
$f(\omega) \propto (1/\omega)\cdot\arctan(\omega \tau_{max}) \propto
1/\omega$, which in Eq.  1 yields a temperature independent relaxation
with $T_{1}^{-1} \propto 1/H$.  This behavior, as observed in our data
above the crossover temperature $T_{0}$ and $H > 1$ T, has previously
been reported for doped semiconductors not far from the
IMT\cite{Fuller96,Paalanen85}.
 
The different behavior of $T_{1}^{-1}(T,H)$ below
$T_{0}$ may be interpreted as evidence for changes in
$f(\omega)$ at low temperatures.  In particular our data for $T \leq
T_{0}$ hints for a rapid increase of the minimum correlation time
$\tau_{min}(T) \propto T^{-2}$ to values much larger than
$\omega^{-1}$ and a drastic reduction of the range of active $\tau$'s. 
This yields $f(\omega) \propto 1/(\omega^{2}\tau_{min})$ and
$T_{1}^{-1} \propto 1/(H^{2}T^{-2})$.

In summary, we have shown that the $^{11}$B-NMR spin-lattice 
relaxation for SrB$_{6}$ and Ca$_{0.995}$La$_{0.005}$B$_{6}$ is 
anomalous.  In case of a large band gap at the $X$ point of the 
Brillouin zone\cite{Tromp2000} our $T_{1}^{-1}$ data can be explained 
by postulating the coexistence of localized and extended electronic 
states with a very weak interaction between them.  With this scenario 
we still face the problem concerning the origin of the weak but 
stable ferromagnetism in these materials. If the band gap is 
small, $T_{1}^{-1}(T)$ might be caused by spin-flips of ``excitonic 
molecules'' pinned to impurity or defect sites.  Finally, we see 
no simple way to explain our $T_{1}^{-1}(T)$ data if there is a band 
overlap.

This work was financially supported by the Schweizerische 
Nationalfonds zur F\"{o}rderung der Wissenschaftlichen Forschung.  We 
greatly acknowledge helpful discussions with R. Monnier, L. Degiorgi 
and M. Zhitomirsky.




\begin{thebibliography}{99}
\bibitem{Spindiffusion} Spin diffusion plays an important role in the
NMR response of doped semiconductors far on the insulating side of the
IMT. In our case, however, the measurements of $m(T)$ at the central
line and at the wings yield the same value for $T_{1}$.  Since the
nuclei at the wings (but not the nuclei at the central line) are
strongly detuned due to the quadrupolar perturbation, spin diffusion
would play a much less important role for the wings than for the
central line.  The identical results for $T_{1}$ in both cases
indicates that spin diffusion effects are negligible.
\end{thebibliography}
\end{document}